\documentclass[aps, superscriptaddress, floatfix, notitlepage, twocolumn, nofootinbib]{revtex4-2}
\usepackage{epsfig}
\usepackage{graphicx}
\usepackage{amsmath}
\usepackage{bbold}
\usepackage{tabularx, booktabs}
\usepackage[normalem]{ulem}
\usepackage{xcolor}
\usepackage{multirow}
\usepackage{dutchcal} 
\usepackage[T1]{fontenc}
\usepackage{longtable}
\usepackage{appendix}
\usepackage[colorlinks=true,allcolors=black]{hyperref}
\usepackage{chemformula}
\let\ce\ch
\usepackage{orcidlink}
\usepackage{lineno}
\usepackage{xr}
\usepackage{float}
\usepackage[caption = false]{subfig}
\newcolumntype{Y}{>{\centering\arraybackslash}X}
\newcolumntype{Z}{>{\hsize=1.1\hsize\centering\arraybackslash}X}

\newcommand{\Tc}{\ensuremath{T_\textrm{c}}}

\newcommand{\Tcad}{\ensuremath{T_\textrm{c}^\text{AD}}}
\newcommand{\Tcalignn}{\ensuremath{T_\textrm{c}^\textrm{pred}}}

\newcommand{\olog}{\ensuremath{\omega_\text{log}}}

\begin{document}
\newcommand{\bochum}{Research Center Future Energy Materials and Systems of the University Alliance Ruhr and Interdisciplinary Centre for Advanced Materials Simulation, Ruhr University Bochum, Universitätsstraße 150, D-44801 Bochum, Germany}
\newcommand{\coimbra}{CFisUC, Department of Physics, University of Coimbra, Rua Larga, 3004-516 Coimbra, Portugal}
\newcommand{\mpi}{Max-Planck-Institut f\"ur Mikrostrukturphysik, Weinberg 2, D-06120 Halle, Germany}
\newcommand{\mlu}{Institut f\"ur Physik, Martin-Luther-Universit\"at Halle-Wittenberg, D-06099 Halle, Germany}
\newcommand{\upv}{Fisika Aplikatua Saila, Gipuzkoako Ingeniaritza Eskola, University of the Basque Country (UPV/EHU), Europa Plaza 1, 20018 Donostia/San Sebastián, Spain}
\newcommand{\cfm}{Centro de Física de Materiales (CFM-MPC), CSIC-UPV/EHU, Manuel de Lardizabal Pasealekua 5, 20018 Donostia/San Sebastián, Spain}
\newcommand{\dipc}{Donostia International Physics Center (DIPC), Manuel de Lardizabal Pasealekua 4, 20018 Donostia/San Sebastián, Spain}
\newcommand{\plb}{Periodic Labs, San Francisco, CA, USA.}

\title{
Search for thermodynamically stable ambient-pressure superconducting hydrides in GNoME database
}

\author{Antonio Sanna$^{\S}$\orcidlink{0000-0001-6114-9552}}
\email{sanna@mpi-halle.mpg.de}
\renewcommand*{\thefootnote}{\fnsymbol{footnote}}
\footnotetext[4]{These authors contributed equally.}
\affiliation{\mpi}\affiliation{\mlu}
\author{Tiago F. T. Cerqueira$^{\S}$\orcidlink{0000-0002-4147-8129}}
\affiliation{\coimbra}
\author{Ekin Dogus Cubuk}
\affiliation{\plb}
\author{Ion Errea\orcidlink{0000-0002-5719-6580}}
\email{ion.errea@ehu.eus}
\affiliation{\upv}
\affiliation{\cfm}
\affiliation{\dipc}
\author{Yue-Wen Fang\orcidlink{0000-0003-3674-7352}}
\email{yuewen.fang@ehu.eus}
\affiliation{\cfm}

\begin{abstract}
\vspace{0.5cm}
Hydrides are considered to be one of the most promising families of compounds for achieving high temperature superconductivity. However, there are very few experimental reports of ambient-pressure hydride superconductivity, and the superconducting critical temperatures (\Tc) are typically less than 10 K. At the same time several hydrides have been predicted to exhibit superconductivity around 100 K at ambient pressure but in thermodynamically 
unfavorable phases.
In this work we aim at assessing the superconducting properties of thermodynamically stable hydride superconductors at room pressure by investigating the GNoME material database, which has been recently released and includes thousands of hydrides thermodynamically stable at 0K.
To scan this large material space we have adopted a multi stage approach which combines machine learning for a fast initial evaluation and cutting edge ab initio methods to obtain a reliable estimation of \Tc. 
Ultimately we have identified 25 cubic hydrides with \Tc~above 4.2~K and reach a maximum  \Tc~of 17 K.  
While these critical temperatures are modest in comparison to some recent predictions, the systems where they are found, being stable, are likely to be experimentally accessible and of potential technological relevance.
\end{abstract}
\maketitle
\newpage

\section{Introduction}
The near room temperature superconductivity, experimentally observed in several superconducting hydrides such as LaH$_{10}$~\cite{Drozdov2019-La-H-Nature,LaH10-PRL2019-Somayazulu} and YH$_9$~\cite{Kong2021-NC-Y-H}  hosting superconductivity above -30 and -23$^{\circ}$C, respectively, demonstrates that the investigation of phonon-mediated superconductors involving the lightest element H is well merited~\cite{FloresBoeri_PerspectiveOnConvetionalHiTcSc_PhysRep2020}. However, achieving near room temperature superconductivity in these hydrides requires subjecting them to extreme pressures over 170 billion pascals (170 GPa), making the real-world application unlikely.

Given the tough conditions of these high-\Tc~superconducting hydrides posed by the high pressures, one may naturally wonder whether it is possible to achieve superconductivity in the hydrides at ambient pressure\cite{roadmap-JPCM}.
In earlier studies, very few hydride superconductors have been experimentally reported at ambient pressure and their superconducting transition temperatures are typically less than 10 K~\cite{JPCM-TiH0.71-1993,PhysRevLett.25.741-1970-Th4H15,NbH0.7-ZPhysikB-1977}. 
The desire to discover high-\Tc\ ambient-pressure superconducting hydrides has driven some ab initio theoretical predictions recently~\cite{Cerqueira-AFM-hydrides2024,sanna2023-Mg2IrH6,dolui2023-Mg2IrH6-PRL2024,Tian2023,AlH4-PRB2023-JunjieSHI,Mg2IrH6-MaterTodayPhy2024-YangSUN,GaoMaxTcArXiv2025,MH4-130K-Wenwen2025}.
In particular, Cerqueira et al combined the machine learning methods with the ab initio study to predict around 50 superconducting hydrides with \Tc~above 20 K from the screening of more than 1 million compounds~\cite{Cerqueira-AFM-hydrides2024}. The cubic hydride family with $M$H$_6$ octahedra is exceptionally attractive because some of them are near the convex hull and can maintain a high \Tc~above the boiling point of liquid nitrogen of 77 K~\cite{sanna2023-Mg2IrH6,dolui2023-Mg2IrH6-PRL2024}. Several perovskite hydrides have also attracted some attention due to their high \Tc~and small enthalpy distance from the convex hull~\cite{Cerqueira-AFM-hydrides2024,perovskite-TianCUI2024AdvSci,perovskite-JunjieSHI-JPCM2024}. Remarkably, ab initio calculations suggest that RbPH$_3$ is dynamically and kinetically stable at ambient pressure with a superconducting critical temperature of about 100 K~\cite{RbPH3-arxiv2024-Dangic}.

As discussed in Refs.~\cite{Cerqueira_Sampling_arxiv2023} and \cite{GaoMaxTcArXiv2025}, the evidence that emerges from recent high-throughput research is that high-\Tc~values of the order of 100~K appear to be systematically linked to some degree of thermodynamical instability.
This may be attributed to the destabilizing effect of delocalized hydrogen-derived electrons in the metallic bonding environment of multinary hydrides. Consequently, the hydrides thermodynamically stable at 0K in large materials database such as Materials Project are generally semiconductors or insulators. 
Clearly this drops a large fraction of the problem to the synthesis, as one expects that a synthesis route for metastable compounds might be very hard to find, 
although some recent advances in pressure-quench protocols suggest that it may be possible to retain superconductivity in the ambient-pressure metastable phases quenched from a pressure where the corresponding phases are thermodynamically stable~\cite{PNAS-ChingwuCHU-FeSe-pressure-quench}.

GNoME has used active learning and graph neural networks to identify 381 thousand crystals that are stable at 0K according to DFT~\cite{AmilMerchant2023Nature-GNoME}. Most of these structures are identified by elemental substitutions, a commonly used method to search for 0K stable crystals~\cite{wang2021predicting,Kirklin2015-OQMD-npjcpmpumats,chen2022universal}. As mentioned in the conclusion of GNoME, these calculations are not able to consider configurational disorder. With further improvements in force fields and computational resources, simulating finite temperature stability in a high throughput manner might be possible. Given this set of crystals that are stable at 0K according to DFT, might there be interesting superconducting candidates? In this work we investigate this possibility by means of first principle simulations.

A brute force use of ab initio methods~\cite{PellegriniReview2024} would be computationally too expensive to be directly applicable to such a large space of materials. For this reason we adopt a recently developed machine-learning accelerated ab intio algorithm. 
This allows to identify possible high-\Tc\ hydrides in the GNoME database and use high level ab initio methods to validate  machine learning predictions.  As we will report below, the search has been successful in identifying 25 cubic hydrides with \Tc~exceeding the boiling point of liquid helium (4.2 K). Most of these hydrides have \Tc\ lower than NbTi alloys (10 K), nevertheless the cubic \ce{LiZrH6Ru} hydride with vacancy-ordered double perovskite structure stands out and shows a \Tc~at 17 K at ambient pressure. 
 
\section{Results}
\subsection{Machine learning accelerated high-throughput ab initio approach}

\begin{figure}[htb]
\includegraphics[angle=0,width=0.49\textwidth]{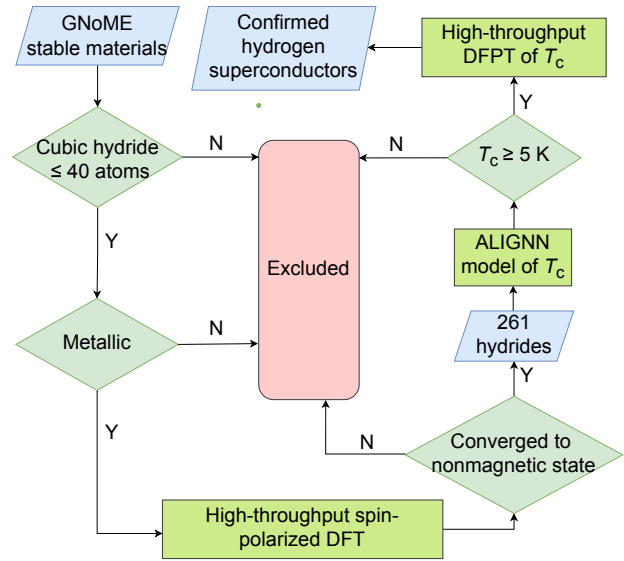}
\caption{The computational workflow of the screening of cubic superconducting hydrides with thermodynamic stability by combining the ab initio study with ALIGNN model. The integers in parentheses indicate the number of hydrides. }
\label{fig:workflow}
\end{figure}

To accelerate the discovery of superconducting hydrides, we have combined the machine learning methods with high-throughput ab initio calculations. The major workflow of the superconductors screening is described in Fig.~\ref{fig:workflow}.

We first performed data mining on thermodynamically stable hydrogen-based compounds at 0K with a cubic structure and metallic properties from the openly available dataset of GNoME project~\cite{gnome}, which resulted in a subset including 490 hydrides.
We focused exclusively on cubic hydrides because the cubic structure is a recurring and well-established structural motif across many classes of high-$T_{\rm c}$ hydrogen-based compounds at both high pressure (e.g. LaH$_{10}$~\cite{Drozdov2019-La-H-Nature,LaH10-PRL2019-Somayazulu} and YH$_9$~\cite{Kong2021-NC-Y-H}) and ambient pressure (e.g. RbPH$_3$~\cite{RbPH3-arxiv2024-Dangic} and Li$_2$AgH$_6$\cite{GaoMaxTcArXiv2025}). 
High-throughput spin-polarized density functional theory (DFT) calculations were further used to find 261 nonmagnetic metallic hydrides. 
Then, the Atomistic Line Graph Neural Network (ALIGNN) model~\cite{Choudhary2021}, which has been proven successful in the discovery of superconductors~\cite{sanna2023-Mg2IrH6,Cerqueira-AFM-hydrides2024}, was used to estimate $T_{\rm c}$ of the hydrides. 
To narrow-down the search space of superconductors and reduce the computational cost of ab initio calculations, only the superconducting hydrides with ALIGNN-predicted \Tcalignn~$\geq$ 1 K were considered in the subsequent high-throughput density functional perturbation theory (DFPT) calculations.

Using high-throughput DFPT methods coupled with Allen-Dynes formula, we found 25 cubic hydrides from the GNoME database with a superconducting critical temperature \Tcad~$\geq$ 4.2 K (see Tables~\ref{tab:double_perovskite_table}-\ref{tab:fluorite_table}). 
For the 25 cubic hydrides, the DFPT-calculated \Tcad~and ALIGNN-predicted \Tcalignn~are compared in a parity plot included in Supplementary Fig. 1. The values of ALIGNN-predicted \Tcalignn~are also included in Tables~\ref{tab:double_perovskite_table} and~\ref{tab:fluorite_table} explicitly. The ALIGNN predicted \Tc~show a strong quantitative agreement with DFPT-calculated \Tcad, with a mean absolute error (MAE) of only 2.5 K.
Given that the MAE of 2.5 K is significantly smaller than the target benchmark (the boiling point of liquid helium, 4.2 K), this predictive uncertainty is acceptable for pre-screening purposes. In addition, the conservative criterion of ALIGNN-predicted \Tcalignn~$\geq$ 1 K should ensure that all materials with true superconducting critical temperature above 4.2 K are retained in the high-throughput DFPT calculations.

By examining the crystal structures of the 25 thermodynamically stable superconducting hydrides, we find that they can be grouped into two prototype structure families, i.e. vacancy-ordered double perovskites and fluorite-like structure with or without vacancy-ordering. 

All the vacancy-ordered double perovskites can be generalized as defected forms of doubly cation ordered ${AA'BB'X_6}$ (see Fig.~\ref{fig:struct_family}(a)). In the defect-free ${AA'BB'X_6}$, $X$ is stabilized at 24f Wyckoff position;
$A$ and $A'$ are located at 4c and 4d Wyckoff positions, and are bonded to twelve equivalent $X$ to form $AX_{12}$ and $A'X_{12}$ cuboctahedra; $B$ and $B'$ reside at 4a and 4b Wyckoff positions, and each $B$ or $B'$ is bonded to six equivalent $X$ ions to form $BX_{6}$ and $B'X_{6}$ octahedra. In the hydrides with vacancy-ordered double perovskite structure, the number of atomic sites is reduced to nine per formula unit due to the existence of vacancies at either 4a or 4b Wyckoff positions.  Fig.~\ref{fig:struct_family}(a) displays \ce{EuDyReH6} as an example of a vacancy-ordered double perovskite where the vacancy appears at the 4a position and hydrogen atoms are located at $X$ site.
In contrast to vacancy-ordered double perovskites with $X$ being H and space group being ${F\bar{4}3m}$ (No. 216), the hydrogen atoms are swapped with the metallic Rh atoms in \ce{Ce2HRh6}, forming the inverted double perovskite with space group of ${Fm\bar{3}m}$ (No. 225), where 4c and 4d positions are both occupied by Ce and the vacancy is located at the 4b position. 
As seen in Table~\ref{tab:double_perovskite_table}, all these vacancy-ordered double perovskites exhibit a superconducting critical temperature ranging from 4.9 K in \ce{ EuDyReH6} to 23.5 K in \ce{LiZrH6Ru}, and most of them show weak electron-phonon ocupling constant $\lambda$ of less than 1. In addition, these materials show a wide range of ~\olog\ from 35.84 K in \ce{EuDyReH6} to 502.93 K in \ce{EuCdFeH6}. The hydride \ce{EuCdH6Ru} has the largest $\lambda$ of 2.53, but its critical temperature remains as small as 13.6 K due to the very small ~\olog\ of 84.40 K.  

Except for vacancy-ordered double perovskites, the remaining materials are fluorite-like (CaF$_2$-like) structured. For example, in the ${Pm\bar{3}m}$ \ce{Ta3NbH8} with 12 atomic sites in the primitive cell, each Ta at 4c position and each Nb at 1a position are bonded in a body-centered cubic geometry to eight equivalent H atoms at the 8c site. In contrast, in some other fluorite-like hydrides such as ${Fm\bar{3}m}$ \ce{Ta6NbH16}, the alternative appearance of vacancy and Nb atoms are observed at the 1a site, leading to 23 atomic sites in the primitive cell. We can find from Table~\ref{tab:fluorite_table} these fluorite-like hydrides show very similar $\lambda$ and  \olog\, with averages of 0.68 and 182 K. As a result, the calculated~\Tcad are very close, all around 5.0-7.0 K.
It is noted that ${P\bar{4}3m}$ LuNH also features a fluorite-like crystal structure with \Tc~of around 16 K and $\lambda$ of around 0.78 at ambient pressure despite that this phase is very thermodynamically unstable--around 0.5 eV/atom above the convex hull~\cite{Zurek2023-LuNH-arXiv2303.15622,2023arxiv2304.04447-Lilia-Chris}. 

\begin{table}[!ht] 
    \caption{Superconducting transition temperatures calculated with the Allen-Dynes formula (\Tcad\ in K) above 4.2 K, logarithmic average of the phonon frequencies (\olog\ in K), and the electron-phonon coupling constant ($\lambda$) for vacancy-ordered double perovskites. 
    The ALIGNN-predicted \Tcalignn~(K)~are included in the last column for comparison with \Tcad.
    }
    \label{tab:double_perovskite_table}
\begin{tabular}{llrrrr}
\toprule
mat\_id\hspace{0.75cm}~ & formula\hspace{0.5cm}~ & $\lambda$\hspace{0.2cm} & $\omega_{\log}$\hspace{0.2cm} & \Tcad\hspace{0.2cm} & \Tcalignn \\
\midrule
  \multicolumn{6}{c}{\bf Vacancy-ordered Double Perovskites} \\
a1c4087396 & \ce{EuDyReH6} & 2.05 & 35.84 & 4.9 & 2.5 \\
2523d9ecde & \ce{NdEuTcH6} & 0.66 & 166.66 & 5.1 & 3.9 \\
1243cdd0e1 & \ce{Ce2HRh6} & 0.67 & 167.00 & 5.1 & 1.6 \\
97ff896f6f & \ce{SmEuTcH6} & 0.67 & 168.28 & 5.3 & 6.2 \\
0eddd1b470 & \ce{EuCdFeH6} & 0.47 & 502.93 & 5.4 & 4.3 \\
cc52a3846d & \ce{EuYTcH6} & 0.67 & 179.97 & 5.6 & 5.7 \\
e5427bdb32 & \ce{TbEuTcH6} & 0.73 & 168.49 & 6.3 & 6.8 \\
84f266aecd & \ce{EuDyTcH6} & 0.74 & 166.47 & 6.5 & 7.7 \\
9248b5ab15 & \ce{EuHoTcH6} & 0.77 & 163.21 & 6.9 & 9.1 \\
4bb5b78a22 & \ce{SmEuReH6} & 0.98 & 105.75 & 6.9 & 2.9 \\
68f1be98a3 & \ce{LiCeMnH6} & 0.65 & 234.44 & 6.9 & 2.7 \\
5e1177e128 & \ce{EuErTcH6} & 0.81 & 160.52 & 7.6 & 9.5 \\
8d4721d779 & \ce{EuYReH6} & 1.28 & 84.82 & 7.7 & 3.3 \\
f68afdcbbc & \ce{EuTmTcH6} & 0.88 & 155.98 & 8.4 & 9.3 \\
1e099eabf5 & \ce{YbCeTcH6} & 0.90 & 150.52 & 8.6 & 6.1 \\
1bd4b03f35 & \ce{YbGdTcH6} & 1.00 & 144.04 & 9.6 & 10.2 \\
8bf27d9be2 & \ce{LiTiH6Ru} & 1.12 & 123.38 & 9.8 & 8.7 \\
e5c2a73dc0 & \ce{EuLuTcH6} & 1.09 & 134.38 & 10.1 & 9.2 \\
51a97ab68d & \ce{EuCdH6Ru} & 2.53 & 84.40 & 13.6 & 6.7 \\
0f97a7734c & \ce{LiZrH6Ru} & 1.00 & 341.59 & 23.5 & 4.7 \\
\bottomrule
\end{tabular}
\end{table}

\begin{table}[htb]
    \caption{Superconducting transition temperatures calculated with the Allen-Dynes formula (\Tcad\ in K) above 4.2 K, logarithmic average of the phonon frequencies (\olog\ in K), and the electron-phonon coupling constant ($\lambda$) for fluorite-like hydrides and vacancy-ordered fluorite-like hydrides.
    The ALIGNN-predicted \Tcalignn~(K)~are included in the last column for comparison with \Tcad.
    }
    \label{tab:fluorite_table}
\begin{tabular}{llrrrr}
\toprule
mat\_id \hspace{0.75cm}~&formula \hspace{0.75cm}~& ~\hspace{0.5cm}$\lambda$\hspace{0.2cm}~&~\hspace{0.5cm} \olog\ \hspace{0.2cm}~& \Tcad & \Tcalignn \\
\midrule
  \multicolumn{6}{c}{\bf Fluorite-like} \\
d8e8825efe & \ce{Ta3VH8} & 0.64 & 178.58 & 4.9 & 4.1 \\
149cfed18b & \ce{Ta3NbH8} & 0.63 & 190.37 & 5.0 & 4.4 \\
34fc41965b & \ce{TaNb3H8} & 0.72 & 185.95 & 7.0 & 7.1 \\
\multicolumn{6}{c}{\bf Vacancy-ordered fluorite-like} \\
140fb04a4b & \ce{Ta6NbH16} & 0.64 & 179.61 & 5.0 & 6.0 \\
d93c6d6bda & \ce{Ta6MoH16} & 0.75 & 176.75 & 7.3 & 6.0 \\
\bottomrule
\end{tabular}
\end{table}

\begin{figure}[htb]
\includegraphics[angle=0,width=0.8\columnwidth]{imag-main/structure-family-v2.pdf}
\caption{The conventional cells of the hydrides selected from vacancy-ordered double perovskites and fluorite-like structures. (a) Normal double perovskite \ce{EuDyReH6} with vacancy ordering  (b) Inverted double perovskite \ce{Ce2HRh6} with vacancy ordering.
(c) Fluorite-like \ce{Ta3NbH8}. (d) Fluorite-like \ce{Ta6NbH16} with vacancy.
The gold spheres represent the ordered vacancies $V_{\rm vac}$. 
}
\label{fig:struct_family}
\end{figure}

\subsection{Improved analysis of \ce{LiZrH6Ru} using McMillan-Allen-Dynes formulas}
\begin{figure}[htb]
\includegraphics[angle=0,width=0.49\textwidth]{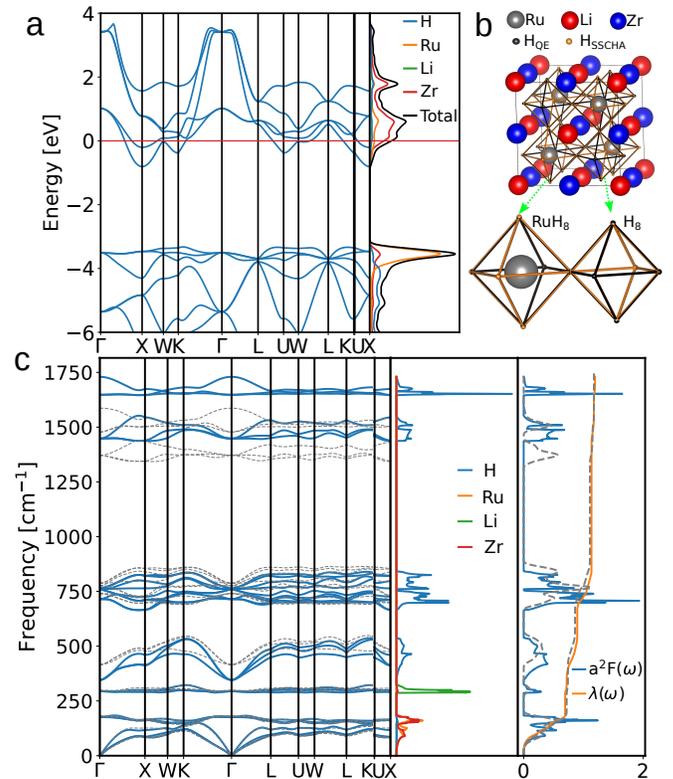}
\caption{The (a) electronic structure, (b) crystal structure, and (c) phonon and electron-phonon coupling properties of \ce{LiZrH6Ru} at ambient pressure. The gray dashed lines in the diagrams of phonon spectrum and Eliashberg spectral functions were calculated by including the ionic quantum and anharmonic effect in SSCHA.}
\label{fig:LiZrH6Ru-0GPa}
\end{figure}

The quarterly hydride  \ce{LiZrH6Ru} has the highest \Tcad~among the studied materials. As seen in Fig.~\ref{fig:LiZrH6Ru-0GPa}(a), there are two bands across the Fermi level. The Li atom is fully ionized and does not contribute to the density of states (DOS) at the Fermi level. Alternatively, the Zr atom dominates the DOS at the Fermi level, with the Ru and H atoms making minor contributions. 

Because the Fermi level is located on a shoulder near the peak of the density of states, we performed the improved accuracy calculations and tested the convergence against the smearing, $k$-grid, and $q$-grid. In the improved calculations with a $k$-grid ($q$-grid) of ${42^3}$ (${6^3}$), the converged $\lambda$ are \olog~are 1.20 and 343.32 K, respectively. Compared to the results of the high-throughput calculations shown in Table~\ref{tab:double_perovskite_table} (i.e. $\lambda$ $\sim$ 1.0 and \olog $\sim$ 341.59 K), the electron-phonon coupling constant and the logarithmic average of the phonon frequencies both increase in accurate calculations and result in a higher \Tcad~of 30.70 K.

Since the ionic quantum and anharmonic effects could be crucial to the crystal structure, dynamical stability and the superconducting properties~\cite{Errea2020-LaH10Nature}, we have further fully relaxed the crystal structure with the stochastic self-consistent harmonic approximation (SSCHA) method~\cite{Errea-PRB2014-SSCHA, Monacelli-JPCM2021-SSCHA, Bianco-PRB2017-SSCHA,monacelli2018pressure}. Compared to the structure obtained in the standard DFT of Quantum Espresso, the lattice constants are unchanged ($a$ = 4.604 \AA, $\alpha$ = 60$^\circ$ in the primitive cell), and the atomic positions of the metal ions are preserved by the symmetry. However, the hydrogen positions are further optimized in SSCHA due to the ioinc quantum and anharmonic effect. 
Because only the hydrogen positions are slightly different in the two crystal structures, we only show the fully relaxed crystal structure from SSCHA calculations in Fig.~\ref{fig:LiZrH6Ru-0GPa}b. As clearly seen by the zoom-in inset in Fig.~\ref{fig:LiZrH6Ru-0GPa}b, there are two different types of octahedra, i.e., RuH$_8$ and H$_8$. 
The volume of  RuH$_8$ octahedron is larger than that of H$_8$ octahedron because the central Ru atom in RuH$_8$ octahedron has a large atomic radius, whereas the H$_8$ octahedron contains a vacant site at its center.
Compared to the crystal structure in harmonic approximation, although every hydrogen octahedron remains perfect without showing any distortion in SSCHA structure, the edges in H$_8$ are slightly decreased from 2.178 \AA~in harmonic approximation to 2.160 \AA~in SSCHA. On the other hand, the edges in RuH$_8$ are slightly increased from 2.426 \AA~in harmonic approximation to 2.444 \AA~in SSCHA.
The optimizations of the hydrogen atoms in the H$_8$ and RuH$_8$ octahedra are also implied in the phonon dispersions in Fig.~\ref{fig:LiZrH6Ru-0GPa}c, in which the whole high-frequency region of the pure hydrogen modes above 1250 cm$^{-1}$ are softened by the anharmonicity renormalization.
Despite the large down shift in the phonon energy, the accumulated electron-phonon coupling is not significantly changed according to the diagram of Eliashberg spectral functions in Fig.~\ref{fig:LiZrH6Ru-0GPa}c. Consequently, there is only a mild improvement in \Tcad, being 32.0 K with quantum effect.

We also studied the effect of pressure on \ce{LiZrH6Ru} in the range from 20 to 200 GPa. \ce{LiZrH6Ru} is found to be dynamically stable up to 180 GPa in the harmonic approximation  and becomes dynamically unstable at 200 GPa due to the appearance of imaginary mode at L point (see Supplementary Fig. 2). 
Our electron-phonon coupling calculations show that the superconducting temperature~\Tcad~can be enhanced at pressures up to 180 GPa compared to that at ambient pressure. In particular, the \Tcad~is boosted to 44 K at 80 GPa. A more specific evolution of \Tcad~and $\lambda$ up to 180 GPa can be found in the Supplementary Fig. 3.
In Fig.~\ref{fig:LiZrH6Ru-pressure-dispersion}, we show the electronic, phonon, and electron-phonon coupling properties at 80 and 160 GPa. 
By comparing the electronic bands at pressures of 80 and 160 GPa with that at ambient pressure (i.e. Fig.~\ref{fig:LiZrH6Ru-0GPa}), the bands become more dispersive with the increase in pressure. This is because the pressure makes the atoms closer, enhancing the  hopping integral between neighboring orbitals and increasing the band width. 
As seen from the DOS diagrams in Fig.~\ref{fig:LiZrH6Ru-0GPa} and Fig.~\ref{fig:LiZrH6Ru-pressure-dispersion}, the DOS at the Fermi level is mostly contributed by Zr at all pressures. we find that only at 80 GPa the Fermi level lies at the sharp peak of projected DOS of Zr while the Fermi level are located at the shoulder of DOS at other pressures. 
In addition, the diagrams of electron-phonon spectral functions $\alpha^2F(\omega)$ (right plots in Fig.~\ref{fig:LiZrH6Ru-0GPa} and Fig.~\ref{fig:LiZrH6Ru-pressure-dispersion}) find that the $\lambda$ value is 1.22 at 80 GPa that is higher than $\lambda$ of 1.19 at 0 GPa and $\lambda$ of 1.11 at 160 GPa.
Consequently, the combination of high DOS at the Fermi level with the large $\lambda$ at 80 GPa is responsible for the highest \Tcad~at this pressure among the studied pressures.

\begin{figure}[htb]
\includegraphics[angle=0,width=0.49\textwidth]{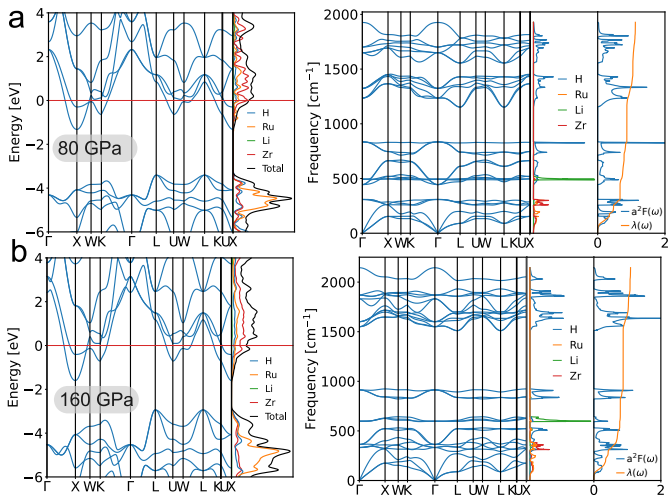}
\caption{Electronic, phonon and electron-phonon coupling properties of \ce{LiZrH6Ru} at (a) 80 and (b) 160 GPa.  
In either panel (a) or (b), the left plot shows the electronic band structure and the atom projected density of states, and the right plot displays the phonon band structure, the atom projected phonon density of states, the electron-phonon spectral function $\alpha^2F(\omega)$ and its integration curve $\lambda(\omega)$.
}
\label{fig:LiZrH6Ru-pressure-dispersion}
\end{figure}

\subsection{In depth characterization of \ce{LiZrH6Ru} beyond McMillan-Allen-Dynes formulas}

So far we have estimated the superconducting properties using the McMillan-Allen-Dynes formulas in the $\mu*$ approximation and assuming a conventional repulsion strength $\mu^*=0.1$. 
This approach is usually reliable when dealing with conventional superconductors~\cite{AllenDynes_PRB1975,PellegriniReview2024}, however, one should consider that it has been developed using data from simple superconducting systems, and it is known to fail in describing systems with complex electronic properties, e.g. Magnesium diboride~\cite{Golubov_MgB2_2002,Mazin_MgB2_2003}.

\ce{LiZrH6Ru} has several features which hint at the possibility that the McMillan   estimation of \Tc\ might not be very reliable. The first thing is that the Fermi level is located close to a peak in the density of states, but close to a band gap. This might lead to a poor Coulomb renormalization~\cite{ScalapinoSchriefferWilkins_StrongCouplingSC_PR1966,Massidda_SUST_CoulombSCDFT_2009,Marini_Chevrel_PRB2021} and an overestimation of \Tc. In simple terms this point can be explained by using the Morel-Anderson formula~\cite{MorelAnderson_1962}, which links $\mu^*$ to the Coulomb repulsion at the Fermi level ($\mu$):
\begin{equation}
\mu^*=\frac{\mu}{1+\mu \ln\left(\frac{E_F}{\omega_{c}}\right)},    
\end{equation}
where $\omega_c$ is a cutoff frequency (arbitrary in Eliashberg theory, close to the maximum frequency of the phonon spectrum in McMillan-Allen-Dynes theory) and $E_F$ is the Fermi energy setting the bandwidth of the valence region.  
In \ce{LiZrH6Ru} the valence band starts about 10 eV below the Fermi level, however the presence of a band gap between -1 and -3 eV complicates the matter. In practice it is not obvious if one should assume $E_F=1$ or $E_F=10$. In fact,  inspecting the character of the electronic bands one can see that bands close to the Fermi level are almost exclusively transition metal states, while low lying states have a dominant H contribution which is small at the Fermi level. 

\begin{figure}[htb]
\includegraphics[angle=0,width=0.8\columnwidth]{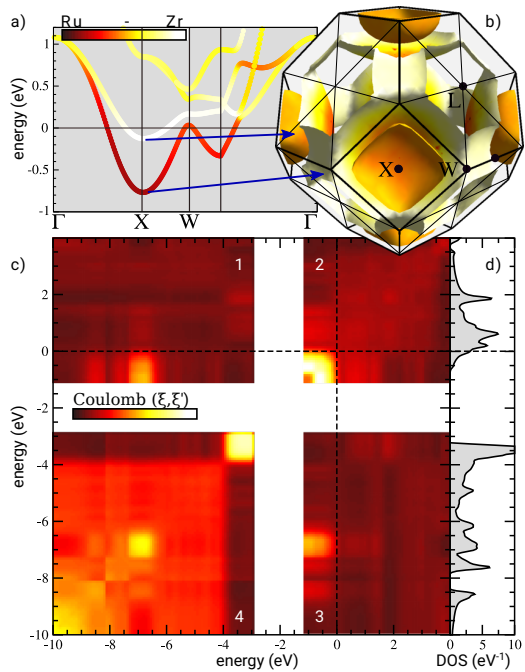}
 \caption{Coupling properties of \ce{LiZrH6Ru}. a) Electronic bands near the Fermi level including  atomic projections on the Zr and Ru sites. b) Fermi surface and the electron-phonon coupling ($\lambda_{\bf k}$). c) Coulomb interaction matrix resolved on iso-energy surfaces. d) Density of electronic states. }
\label{fig:LiZrH6Ru_analysis}  
\end{figure}

Second, we see that the Fermi level states have a strong Ru and a Zr component. It is not uncommon that superconductivity in transition metals is affected by incipient magnetism. For example the critical temperature of \ce{Nb} (right next to \ce{Zr} in the periodic table) is significantly reduced by the effect of spin fluctuations~\cite{Rietschel_SpinFluctVandNb_PRB1979,Kentaro_SpinFluctVandNb_PRB2020,PellegriniKukkonenSanna_beyondRPA_PRB2023}.
Last but not least, as shown in Fig.~\ref{fig:LiZrH6Ru_analysis}a) there are two electronic bands crossing the Fermi level with  different orbital character. Multi-band effect in superconductivity might lead to an increase \Tc\ as compared to a single band approximation~\cite{SuhlMatthiasWalker_multibandBCS_PRL1959,Mazin_MgB2_2003,Floris_MgB2_PRL2005,Floris_MgB2_PhysicaC2007}. The ${\bf k}$-resolved electron-phonon coupling ($\lambda_{\bf k}$) on the Fermi surface is shown~\cite{Kawamura_fermisurfer} in Fig.~\ref{fig:LiZrH6Ru_analysis} showing a substantial coupling anisotropy, although neither strong or simple as in \ce{MgB2}. 

The first points can be addressed by computing the critical temperature by solving the Eliashberg equations including the Coulomb interaction from first principles~\cite{Sanna_Eliashberg_JPSJ2018,Pellegrini_SimpEliashberg_JOPM2022}, this approach includes the full electronic bandwidth of the problem and the dielectric screening in the random phase approximation. The Coulomb matrix is shown in Fig.~\ref{fig:LiZrH6Ru_analysis}. This 2D matrix  ($W(\xi,\xi')$) is defined by averaging the Coulomb matrix elements between iso-energetic electronic states~\cite{Massidda_SUST_CoulombSCDFT_2009,PellegriniKukkonenSanna_beyondRPA_PRB2023}. By definition $\mu=W(0,0)N_F$, where $N_F$ is the DOS at the Fermi level. In the random phase approximation (RPA) we have $\mu=0.66$ which is quite large, but not uncommon in transition metal compounds~\cite{PellegriniKukkonenSanna_beyondRPA_PRB2023}. As one can see this matrix has approximately a block diagonal structure. The main valence states (panel c4) are strongly repulsive in the -10 to -4 eV range, but they interact weakly with the Fermi level states (panels c3 and c1). This fact, the large value of $\mu$ and 
  the presence of a large band gap, necessary lead to an extremely disruptive Coulomb contribution to the superconducting state. The calculated \Tc\  solving the isotropic Eliashberg in RPA is 14.2~K. This estimation is about half of that obtained by the standard McMillan-Allen-Dynes approach, an uncommon deviation which reminds the case of Chevrel phases discussed in Ref.~\cite{Marini_Chevrel_PRB2021}.

  A second aspect we would like to further investigate is the possible role of magnetic effects in superconductivity. \ce{LiZrH6Ru} is non magnetic in the sense that we could not stabilize any ferromagnetic or antiferromagnetic ground state order. However transition metals might have an incipient tendency to magnetism which results in a large spin-susceptibility which enhances the Coulomb repulsion. In extreme cases this might lead to unconventional superconductivity~\cite{Essenberger_SpinFluctuations_PRB2014,Essenberger_IronBasedSC_PRB2016} and in more conventional cases to a reduced value of \Tc ~\cite{Rietschel_SpinFluctVandNb_PRB1979}. This can be included in our ab initio Eliashberg formalism using the Kukkonen-Overhauser approach discussed in Ref.~\cite{PellegriniKukkonenSanna_beyondRPA_PRB2023,Kukkonen_elelInteractionSimpMetals_PRB1979}. This leads to a \Tc\ estimation of 11.3~K. Considering that, the ab initio inclusion of the spin susceptibility tends to give a stronger Coulomb repulsion even in absence of significant magnetic instabilities, which is consistent with a marginal role of magnetic effects. 

The final aspect we wish to address is the possibility of multiband / anisotropic superconductivity. With this motive we have computed the 2-band resolved $\alpha^2F$ function and solved the multiband Eliashberg equation. This approach gives an enhanced \Tc\ of about 20\%. A nearly identical enhancement was computed using the fully anisotropic  superconducting density functional theory (SCDFT) approach~\cite{SPG_EliashbergSCDFT_PRL2020}, proving that the two-band effect is the leading contribution to the anisotropic enhancement of \Tc. As shown in Fig.~\ref{fig:LiZrH6Ru_analysis}a), 
the two portions of the Fermi surface are chemically different and spatially disjoint. 
The inner Fermi surface has a mix Ru-Zr character, while the outer one has a pure Zr projection. 
The components of the electron-phonon coupling from the inner and outer bands are $\lambda\sim0.84$ and $\lambda\sim0.3$ respectively. 
While these values alone could suggest strong anisotropy effects, similar to the case of \ce{MgB_2}, unlike \ce{MgB_2} the inter-band coupling is also very large,  scattering from the outer band to the inner one has $\lambda=0.86$. In this situation the band anisotropy is washed out by the electron-phonon coupling, leading to a minor anisotropy contribution to \Tc. In the language of the Suhl, Matthias, Walker paper~\cite{SuhlMatthiasWalker_multibandBCS_PRL1959} anisotropy has a small effect on \Tc\ because the maximum eigenvalue of the $\lambda$ matrix (1.32) is close to the isotropic coupling (1.26). Our best estimation of \Tc\ including multiband effects and the ab-initio RPA Coulomb interaction is 17~K. 

It is natural to compare this relatively low \Tc\ predicted for \ce{LiZrH6Ru} and those predicted for a similar family of compounds Mg$_2$$M$H$_6$ ($M$ = Rh, Ir, Pd, or Pt) in Ref.~\cite{sanna2023-Mg2IrH6} and Refs. \cite{dolui2023-Mg2IrH6-PRL2024,Mg2IrH6-MaterTodayPhy2024-YangSUN}.  The key difference is that in the Mg$_2$$M$H$_6$ family hydrogenic states are present at the Fermi level and provide a significant contribution to the superconducting coupling. On the other hand in the \ce{LiZrH6Ru} system the hydrogen contribution is negligible on the Fermi surface. A neraly identical situation has been discussed in detail in Ref.~\cite{Zurek_Mg2IrH6_AngewChem2024} to explain the different superconducting properties of \ce{Mg2IrH6} and \ce{Ca2IrH6}. The fact that the Mg$_2$$M$H$_6$ family has high \Tc\ but is thermodynamically unstable while \ce{LiZrH6Ru} is thermodynamically stable, is consistent with the recent analysis reported in Ref.~\cite{GaoMaxTcArXiv2025} about the limits of \Tc\ in conventional superconductors. On a positive note, the synthesis of \ce{LiZrH6Ru} might be easier owing to the favorable thermodynamics, while that of \ce{Mg2IrH6} is expected to be challenging~\cite{Hansen_MgIrH5_pathway_PRB2024}.

\subsection{A note about ab initio predictions and their reliability}

The predictive power of computational methods in superconductivity research has evolved substantially over the past decades; however, several challenging limitations remain that may affect the confidence of ab initio predictions. Therefore, it is important to discuss what we consider to be the main sources of uncertainty that may lead to discrepancies between computational predictions and experimental realizations.

The first major source of uncertainty lies in the identification of thermodynamically stable phases. 
The typical error in energy estimations, while strongly material dependent, is of the order of 50 meV$\cdot$atom$^{-1}$~\cite{Wang2021-scireport}. Therefore, if the energy cost for a thermodynamic decomposition is less than this, theoretical predictions should be considered extremely uncertain.

A second source of uncertainty derives from the fact that, even if a phase is truly thermodynamically stable, its synthesis might not achieve a sufficiently pure form so that predictions for a perfect bulk phase still apply. 
This is mostly a matter of experimental work, when superconductors are of great technological interest (e.g. high temperature superconductors) synthesis methods are investigated with utmost care to achieve samples with the highest quality. In other cases there might not be a sufficient motivation to invest resources in optimizing the synthesis process.
This consideration also applies to metastable phases, which may be accessible only through complex synthesis pathways and are therefore worth investigating primarily when a compound is expected to exhibit exceptional properties. This may be the case for Mg$_2$IrH$_6$, which is predicted to have an extremely high Tc~\cite{sanna2023-Mg2IrH6,Cerqueira-AFM-hydrides2024}; however, its slight metastability led, upon attempted synthesis, to the formation of Mg$_2$IrH$_5$, which is unfortunately a semiconductor~\cite{Hansen_MgIrH5_pathway_PRB2024}.

A third source of uncertainty stems from the theory of superconductivity itself and from the state-of-the-art approximations we employ. A precise error estimate is difficult to obtain; however, unless \Tc\ is very low, modern superconductivity methods typically carry an expected uncertainty on the order of 30\% \cite{PellegriniReview2024}. It should also be noted that some studies focus primarily on structural predictions and do not apply sufficient rigor to the prediction of \Tc.

\section{Conclusion}
We have investigated if within the recently released GNoME dataset of predicted stable (at 0K) compounds, there is any exceptional hydride superconductor with high critical temperature. Our high-throughput calculations was able to identify 25 superconductors with \Tc\ ranging from boiling point of liquid helium (4.2 K) to 23.5 K. For the highest-\Tc\ compound, advanced ab initio methods refined \Tc\ down to 17 K. While not extremely high, as compared to some of the other predictions which have appeared in literature, these stand out because of the extremely accurate estimation of thermodynamic stability at 0K (due to a more densely sampled 0K convex hull). At this stage, an experimental feedback that can validate both the crystal structures and superconductivity of these DFT-based candidates would greatly benefit the future search for superconductors.

\section{Methods}\label{sec:DFT-methods}

All the ab initio calculations were computed by the Vienna ab initio simulation package (VASP)~\cite{Kresse1996,Kresse-PRB-1996} and Quantum Espresso (QE)~\cite{QE-2009,QE-2017}. In both VASP and Quantum Espresso calculations, the PBEsol functional, a revised version of the Perdew-Burke-Ernzerhof (PBE) generalized gradient approximation (GGA) for solids~\cite{PBEsol-Perdew2008}, was employed for exchange-correlation effects in all the ab initio calculations. In the QE calculations, the PBEsol pseudo potentials from the strict set of PseudoDojo were employed~\cite{vanSetten2018pseudodojo}. For elements with $f$ electrons, it is noted that they are frozen as the core states in the PseudoDojo potentials. 

To prescreen the nonmagnetic and metallic hydrides, spin-polarized density functional theory (DFT) methods implemented in the VASP was used and those calculations were managed and run by the Quacc code~\cite{rosen_quacc2024}. The kinetic energy cutoff of 550 eV was used for the plane-wave basis. In the structure optimizations, a grid density of 1000 per Angstrom$^{-3)}$ of reciprocal cell was used to sample the $k$-grid. Electronic self-consistency was achieved with an  energy convergence threshold of 10$^{-6}$ eV.  The relaxations were carried out until the forces on all the atoms were smaller than 0.001 eV/\AA.

We trained the \textsc{alignn}~\cite{Choudhary2021} model using as targets, simultaneously, $\lambda$, \olog\ and \Tc\, with the error for each property weighted equally, as these are the choices yielding the best results. We used the default hyperparameters. Data for the training dataset was obtained from Refs.~\cite{Cerqueira_SamplingConventional_AdvMat24,sanna2023-Mg2IrH6,Cerqueira-AFM-hydrides2024}, and can be download from \url{https://alexandria.icams.rub.de}.
The used training dataset ensures that all elements are included except the lanthanides and actinides, with La and Lu as the only exceptions. It is noted that the \Tc~expands a wide range up to to 80 K, providing some coverage of high-\Tc~regimes. In addition, all crystal systems except monoclinic and triclinic are represented in the training set, including cubic hydride systems which are relevant for our target materials in this study.

The other ab initio calculations were all carried out by using  Quantum ESPRESSO~\cite{QE-2009,QE-2017}. In the high-throughput Quantum Espresso calculations, geometry optimizations employed uniform $\Gamma$-centered $k$-point grids with a density of 3000 $k$-points per reciprocal atom. When this resulted in an odd number of $k$-points in any direction, the next even number was used. Energy, force, and stress convergence thresholds were set to $1\times10^{-8}$~a.u., $1\times10^{-6}$~a.u., and $5\times10^{-2}$~kbar, respectively. For electron-phonon coupling calculations, we utilized a double-grid technique where the $k$-grid from lattice optimization was doubled in each direction for the coarse grid and quadrupled for the fine grid. Phonon $q$-sampling used half of the aforementioned $k$-point grid. The Eliashberg function was obtained through double $\delta$-integration using a Methfessel–Paxton smearing of 0.05~Ry.

Ab initio Eliashberg simulations including Coulomb interactions are performed using the approach of Ref.~\cite{Pellegrini_SimplifiedEliashberg2022}, where Coulomb matrix elements and average Coulomb interaction functions $W(\xi,\xi')$ are computed using the elk code~\cite{ElkCode}. SCDFT anisotropic simulations are done with a Monte-Carlo algorithm using 35 K k-points accumulated with higher probability near the Fermi surface~\cite{Sanna_NbSe2_npjQM2022,PellegriniReview2024}. The same algorithm is used to compute the electron-phonon spectral functions for multiband simulations. Electronic and coupling parameters are linearly interpolated on the random-mesh which is used for the solution of the SCDFT Kohn-Sham gap equation. We adopt the functional from  Ref.~\cite{SPG_EliashbergSCDFT_PRL2020}. 
Fermi surfaces are plotted with the Fermisurfer code~\cite{Kawamura_fermisurfer}.

\section{Code availability}

The first-principles calculations were performed using the proprietary code VASP~\cite{Kresse1996,Kresse-PRB-1996}, and Quantum ESPRESSO~\cite{QE-2009,QE-2017} which is an open-source suite of computational tools with the GNU General Public License v2.0. 
The phonon and electron-phonon properties were calculated by
Quantum ESPRESSO~\cite{QE-2009,QE-2017} and SSCHA~\cite{Errea-PRB2014-SSCHA, Monacelli-JPCM2021-SSCHA, Bianco-PRB2017-SSCHA,monacelli2018pressure}. 
The SSCHA code (\url{https://github.com/SSCHAcode/python-sscha}) is open source and is based on the GNU General Public License v3.0.
Quantum ESPRESSO (https://www.quantum-espresso.org) is an open source suite of computational tools with the GNU General Public License v2.0. 
The crystal structure visualization software
VESTA~\cite{momma2011vesta} is distributed free of charge for academic
users under the VESTA License (\url{https://jp-minerals.org/vesta/en/}).

\section{Data availability}
The data supporting the findings of this study are included within the article and its Supplementary Information.

\section{Acknowledgments}
We are grateful to Miguel A. L. Marques for the fruitful discussions and the development of some computational tools used in our study.
This project is funded by the European Research Council (ERC) under the European Union's Horizon 2020 research and innovation program (Grant Agreement No. 802533) and the Department of Education, Universities and Research of the Eusko Jaurlaritza and the University of the Basque Country UPV/EHU (Grant No. IT1527-22). 
 The authors acknowledge the financial support received from the IKUR Strategy under the collaboration agreement between Ikerbasque Foundation and Centro de Física de Materiales (CFM-MPC) on behalf of the Department of Science, Universities and Innovation of the Basque Government (HPCAI21: AI-CrysPred).
This project is also partially supported by the Extraordinary Grant of CSIC (No. 2025ICT122).
T.F.T.C acknowledges financial support from FCT - Fundação para a Ciência e Tecnologia, I.P. through the project CEECINST/00152/2018/CP1570/CT0006 with
DOI identifier 10.54499/CEECINST/00152/2018/CP1570/\linebreak{}CT0006, and computing resources provided by the project Advanced Computing Project 2023.14294.CPCA.A3, platform Deucalion.
Y.-W.F. and I.E. acknowledge PRACE for awarding access to the EuroHPC supercomputer LUMI located in CSC's data center in Kajaani, Finland through EuroHPC Joint Undertaking (EHPC-REG-2022R03-090 and EHPC-REG-2024R01-084). Technical and human support provided by DIPC Supercomputing Center is gratefully acknowledged.
The authors acknowledge enlightening discussions with the partners of the SuperC collaboration.

\section{Author contributions}
Y.-W.F. conceived the study. Y.-W.F. and I.E. planned the research.
T.F.T.C., A.S., and Y.-W.F. performed the theoretical calculations.
Y.-W.F. and A.S. wrote the manuscript with substantial contributions from all the other authors. 
All authors contributed to the analysis of the results.

\section{Ethics declarations}
\textbf{Competing interests} \\
The authors declare no competing interests.

\section{Additional information}
Materials \& Correspondence should be addressed to A.S., I.E. or Y.-W.F.


%
\end{document}